\documentclass{revtex4}
\usepackage{graphics}
\usepackage{latexsym}
\usepackage{amsmath,amssymb}
\usepackage{epsfig,graphicx}
\def\be{\begin{equation}}
\def\ee{\end{equation}}
\def\bea{\begin{eqnarray}}
\def\eea{\end{eqnarray}}
\begin{document}


\title{Cosmology with Galaxy Correlations}
\author{Robert C. Nichol}
\email{bob.nichol.AT.port.ac.uk}
\affiliation{Institute of Cosmology \& Gravitation, University of
Portsmouth, Portsmouth~PO1~2EG, UK}

\begin{abstract}
In this review, I outline the use of galaxy correlations to constrain cosmological parameters. As with the Cosmic Microwave Background (CMB), the density of dark and baryonic matter imprints important scales on the fluctuations of matter and thus the clustering of galaxies, e.g., the particle horizon at matter--radiation equality and the sound horizon at recombination. Precision measurements of these scales from the Baryon Acoustic Oscillations (BAO) and the large scale shape of the power spectrum of galaxy clustering provide constraints on $\Omega_{m}h^2$. Recent measurements from the Sloan Digital Sky Survey (SDSS) and 2dF Galaxy Redshift Survey (2dFGRS) strongly suggest that $\Omega_{m}<0.3$.  This forms the basic evidence for a flat Universe dominated by a Cosmological Constant ($\Lambda$) today (when combined with results from the CMB and supernova surveys). Further evidence for this cosmological model is provided by the late--time Integrated Sachs--Wolfe (ISW) effect, which has now been detected using a variety of tracers of the large scale structure in the Universe out to redshifts of $z>1$. The ISW effect also provides an opportunity to discriminate between $\Lambda$, dynamical dark energy models and the modification of gravity on large scales.

\end{abstract}

\maketitle

\section{Introduction}
\label{intro}
\begin{figure*}[t]
  \includegraphics[width = 0.75\textwidth]{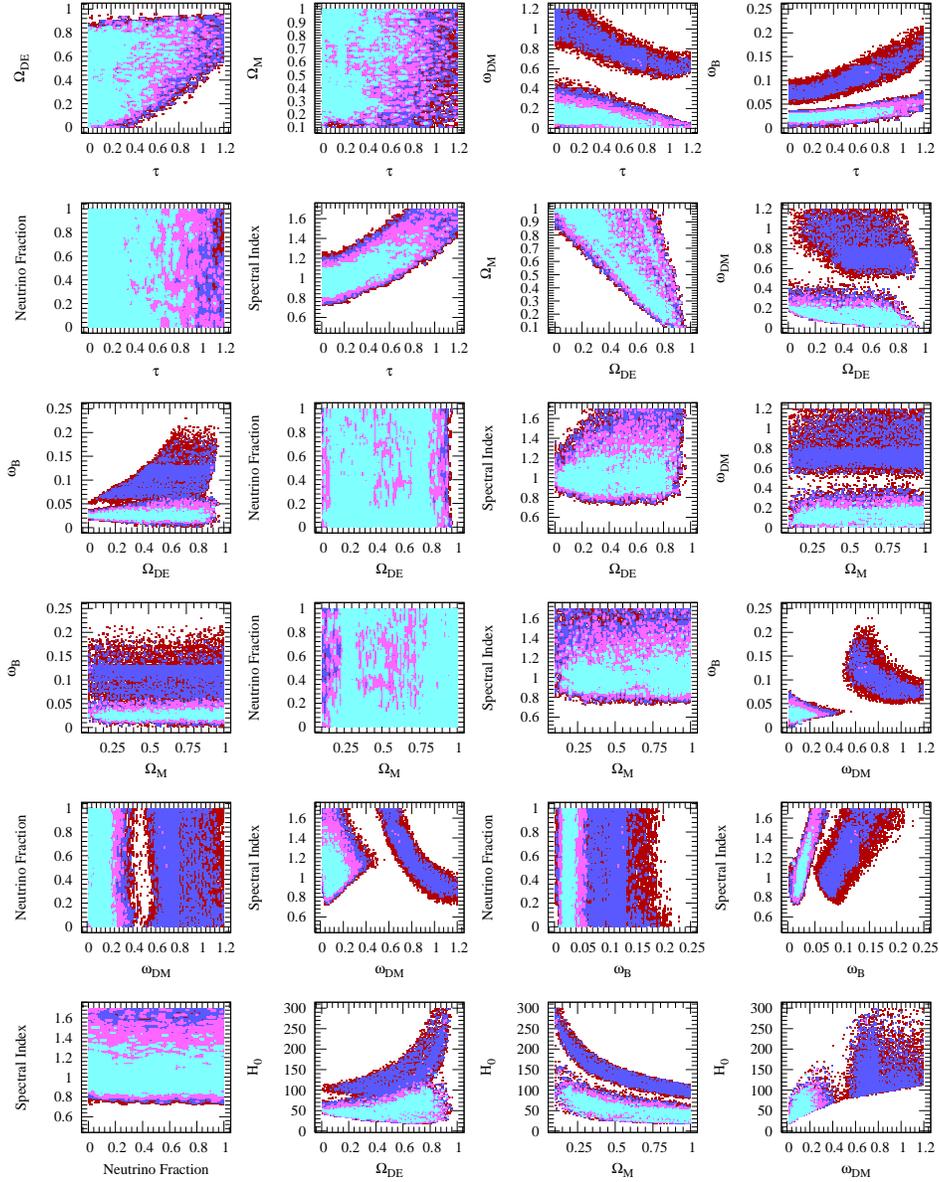}
  \caption{Using only the WMAP data, this figure shows the jointly valid confidence regions for pairs of cosmological
parameters for a 7--dimensional parameter space of $\tau, \Omega_{DE}, \Omega_M, \omega_{DM}, \omega_b, f_{nu}, n_s$, i.e., the optical depth, dark energy mass fraction, total mass fraction, density of dark matter, baryon density, neutrino fraction and spectral index respectively.  The colours cyan, magenta, blue and red correspond to 
$\frac{1}{2} \sigma, \sigma, 1 \frac{1}{2} \sigma$ and $2\sigma$
confidence levels respectively.
Areas of solid color indicate values for the given
pair of fixed (plotted) parameters that contain the true value of the
cosmological parameter, regardless of the
values of the remaining 6  parameters.
Note there are two disjoint regions in parameter space
   which are above the $2\sigma$ confidence interval. Figure is taken from \cite{4}. \label{fig1}
}
\end{figure*}

\begin{figure*}[t]
 \includegraphics[width = 0.75\textwidth]{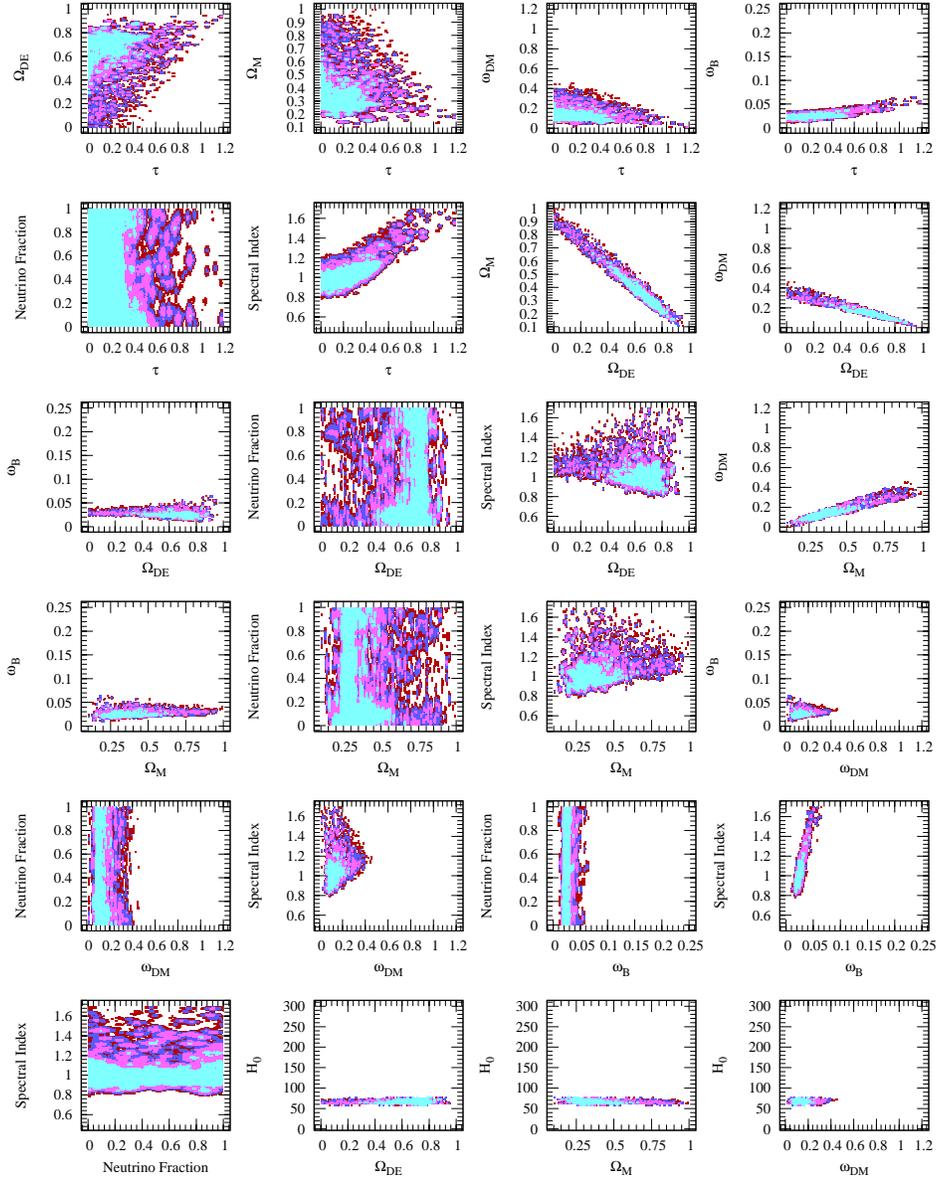}

\caption{The same as Figure \ref{fig1} but now assuming an external constraint on Hubble's Constant of $0.6<h<0.75$. Figure is taken from \cite{4}.}
\label{fig2}
\end{figure*}

In this review, I will focus on the cosmological constraints provided from the correlations of galaxies on large scales in the Universe. Such constraints are complementary to measurements using data from supernovae (reviewed by Bruno Leibundgut in this journal) and the Cosmic Microwave Background (CMB).  I do not attempt to give a complete review of the recent CMB observations here as they have been thoroughly reviewed elsewhere\cite{1}. I just note that the CMB data alone can not simultaneously constrain all the important parameters of the present ``concordance model" of cosmology\cite{2}. This is illustrated in Figure \ref{fig1}, which shows the allowed confidence regions when jointly exploring a seven-dimensional parameter space of $\tau$, $\Omega_{de}$, $\Omega_m$, $\omega_{dm}$, $\omega_b$, $f_{nu}$ and $n_s$ using only the WMAP data. Degeneracies in these parameters can only be broken by adding external data, as demonstrated in Figure \ref{fig2}. For example, there are degeneracies between the properties of dark energy and the curvature of the Universe when only using CMB data. However, if one assumes a Cosmological Constant ($\Lambda$) for the dark energy, then the WMAP data does provide a constraint (95\% confidence) of $-0.08<\Omega_K<0.02$ \cite{3} indicating that the Universe is close to spatially flat (i.e., $\Omega_{total}=\Omega_m+\Omega_{\Lambda}=1$ with $\Omega_K=0$). Therefore, I assume flatness for most of this review unless stated otherwise.

\begin{figure*}[t]
 \includegraphics[width = 0.9\textwidth]{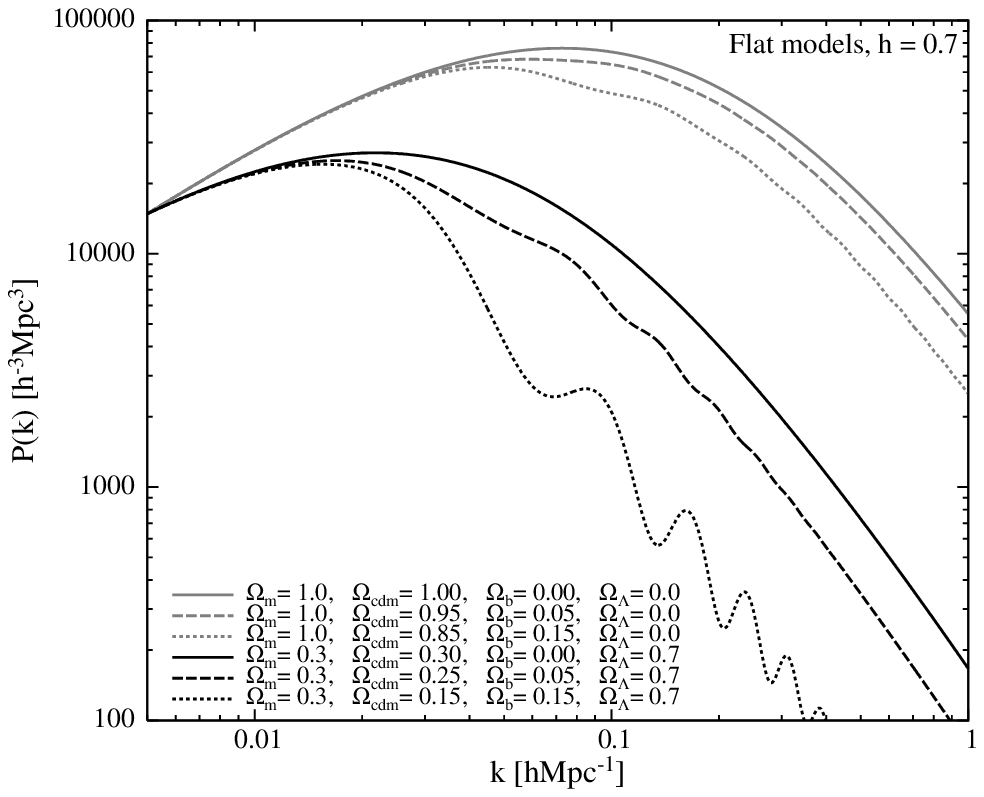}

\caption{The power spectrum of clustering of matter in the Universe as a function of the density of all matter ($\Omega_m$), (cold) dark matter ($\Omega_{cdm}$, dark energy ($\Omega_{\Lambda}$ and baryonic matter ($\Omega_b$). Only flat models are presented here, i.e., all densities must sum to one. Figure kindly provided by Gert Hutsi. A similar plot can be found at \cite{5}.}
\label{hutsi}
\end{figure*}

\section{Clustering of Galaxies}

\subsection{Theoretical Expectations}
\label{theory}

The clustering of galaxies, as a function of scale, contains significant cosmological information. As shown in Figure \ref{hutsi}, the power spectrum of clustering of matter ($P(k)$, see below) depends on both the overall density of dark matter ($\Omega_{cdm}$ in Figure 3) and baryonic matter ($\Omega_b$). In particular, the broad shape of $P(k)$ changes as the ``turnover" from the large scale, scale-invariant spectrum coming out of Inflation is shifted to larger $k$ values (smaller scales) as the density of the total matter decreases ($\Omega_m$). This is caused by the change in the particle horizon scale at the epoch of matter--radiation equality\cite{5}, which is proportional to $\Omega_m h^2$. 

Figure \ref{hutsi} also shows the effect of baryons on the power spectrum. Firstly, baryons decrease the growth rates between the matter--radiation epoch and the drag epoch, leading to a suppression in the overall amplitude of $P(k)$ on small scales relative to the large scales\cite{6}. Secondly, there is an obvious oscillatory signal which increases in amplitude as the baryon density increases. These ``Baryon Acoustic Oscillations" (BAO) are due to sound waves in the early Universe. Before recombination, the intense pressure within the initial dark matter density fluctuations drive sound waves out through the primordial plasma at approximately half the speed of light\cite{7}. These sound waves propagate through the plasma until recombination with the baryons and photons in equilibrium. 

At recombination, the photons and baryons de--couple as the free electrons are captured to form neutral hydrogen. The photons stream through the Universe, rarely interacting with matter again (see Section \ref{isw}) and are seen today as the CMB, redshifted from infra-red wavelengths (one micron) at recombination to radio wavelengths today by the expansion of the Universe. The effect of these sound waves is also seen in the power spectrum of the CMB radiation as the ``acoustic peaks" and provide information on the density of dark and baryonic matter in the Universe (from their relative heights)\cite{1}.

As the photons stream away, there is a drop in the pressure and the baryons are stalled with a radius equal to approximately the sound speed times the age of the universe at recombination (the sound horizon at recombination), which is $\simeq154$ Mpc today\cite{7}. As the Universe evolves after recombination, the baryons in the stalled sound wave move back into equilibrium with the dark matter halos that produced the wave in the first place. However, we still expect evidence for these stalled sound waves to persist today as a preferred scale of clustering in the statistical distribution of galaxies. This is because the overdensities caused by the stalled sound waves help to accelerate galaxy formation leading to a slight excess in the number of galaxy pairs separated by this acoustic scale (in co--moving coordinates).  

I have only considered linear scales here. On smaller scales, the non--linear effects of structure formation will also change the shape of the power spectrum, while other components in the Universe (e.g., relativistic neutrinos) will also leave their mark, see \cite{8} for more details. 

\subsection{Statistical Methods of Galaxy Clustering}
\label{stats}

The most popular statistical method for measuring the clustering of matter in the Universe is the two--point autocorrelation function ($\xi(r)$), which quantifies the probability ($P_{12}$) of finding a pair of galaxies (separated by distance $r$) compared to a random distribution of galaxies, i.e., 

\begin{equation}
P_{12} = \bar{n}^2 (1+\xi(r)) dV_1 dV_2,
\end{equation}

\noindent where $\bar{n}$ is the mean space density of galaxies, and $dV_1$ and $dV_2$ are the search volumes for the two galaxies in question (with subscripts 1 and 2). As one can see, when $\xi(r)=0$ the probability is just that expected from a random distribution of galaxies. If  $\xi(r)>0$, then the galaxies are more clustered than random (higher probability of finding a pair of galaxies separated by $r$), while $\xi(r)<0$ means the galaxies are less clustered than random. 

Practically, $\xi(r)$ is computed by counting all galaxy pairs, as a function of their separation distance, and then comparing this with a similar counting of pairs of random points distributed over the same volume as the real data. This can be written as

\begin{equation}
\xi(r) = \frac{DD(r)}{RR(r)} -1,
\label{eqn2}
\end{equation}

\noindent where $DD(r)$ is the number of data--data pairs and $RR(r)$ is the number of random--random pairs over the same volume as the real data. If $DD=RR$, then $\xi(r)=0$ as expected for a random distribution. There are more optimal estimators of the correlation function\cite{9} than shown in Eqn \ref{eqn2}, which do a better job in accounting for edge effects and the weighting of the pairs. The real complexity comes in the construction of the random data-set for comparison with the real data. This must account for all effects on the real data, including the selection function, which can be challenging\cite{10}. 

In recent years, many authors have preferred to use the power spectrum ($P(k)$) of galaxy fluctuations which is related to $\xi(r)$ by

\begin{equation}
P(k) = \int \xi(r) e^{ik.r} d^3 r,
\end{equation}

\noindent where the power spectrum is just the Fourier transform of the correlation function\cite{10}. Therefore, they both contain the same information and for modern galaxy surveys, with complex masks and selection functions, there is no practical difference between the two methods. 

In reality, one measures the separation of galaxies in ``redshift--space'' meaning the observed redshifts of galaxies (from their spectra) have been converted into radial (line--of--sight) distances and therefore, any intrinsic velocities of the galaxies (in their cosmological rest--frame) will be added to the redshift. This causes the distances to be distorted compared to the ``real--space'' measurements, i.e., with no ``peculiar velocities'' (as astronomers call them) present. Authors often use the symbol $s$, rather than $r$, to signify redshift--space distances, i.e., $\xi(s)$ is the correlation function in redshift--space. On small scales, in the centres of clusters of galaxies, such velocities can be large ($>1000{\rm km\,s^{-1}}$) and can cause noticeable ``Fingers--of--God'' in the distribution of galaxies pointing back to us, the observers. This is because we only see the addition of the radial component of the peculiar velocities to the radial redshift direction of the galaxy survey, leaving the angular coordinates of the galaxies unaffected. 

\begin{figure*}[t]
 \includegraphics[width = 0.9\textwidth]{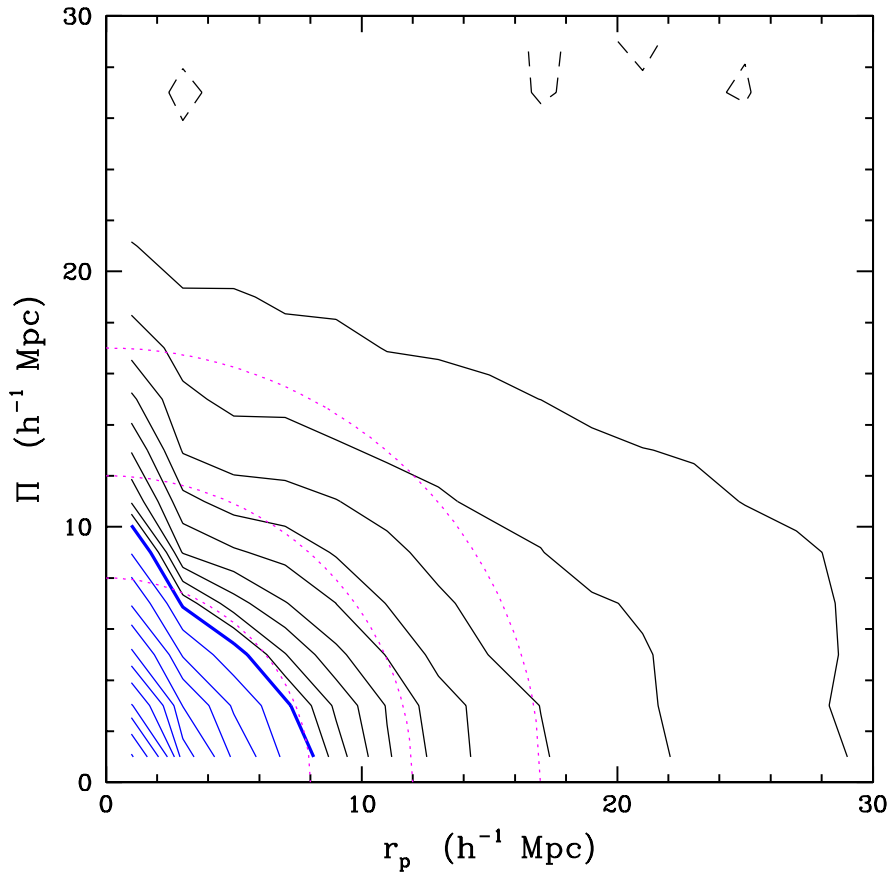}

\caption{The correlation function as a function of the perpendicular ($r_p$) and radial, or line--of--sight, ($\pi$) distances for galaxy pairs from the Sloan Digital Sky Survey main galaxy sample. The thick (blue) contour signifies $\xi(r_p,\pi)=1$, while the dashed (blue) contours are  $\xi(r_p,\pi)=0$.  The dotted circular lines show the expected form of $\xi(r_p,\pi)$ in real-space without any distortions due to galaxy velocities (see text). As can be seen, on small $r_p$ scales, the correlation function is elongated along the $\pi$ axis due to the ``Fingers of God". On large $r_p$ scales, the contours are flattened with respect to the dotted lines due to the coherent infall of galaxies into clusters and superclusters known as the ``Kaiser Effect''. Figure taken from \cite{11}.}
\label{zehavi}
\end{figure*}

This effect can be clearly seen by computing the correlation function as a function of both the tangential, or perpendicular ($r_p$), distance and radial, or parallel ($\pi$), distance between galaxies, i.e., $\xi(r_p,  \pi)$. In this way, one can integrate the correlation function in the radial direction to ``average over" the peculiar velocities thus gaining an estimate of the real--space correlation function, i.e., 

\begin{equation}
w_p(r_p) = 2 \int^{\pi_{limit}}_0 \xi(r_p,  \pi) d\pi,
\end{equation}

\noindent where $\pi_{limit}$ is the upper limit of the integration in the radial direction (the factor of 2 is there because we assume symmetry in the $-\pi$ direction and thus the lower integration limit starts at zero and not $-\pi_{limit}$).  The details of $\pi_{limit}$ are survey dependent, but it is usually tens of megaparsecs\cite{11}. This function is often called the projected correlation function and an example of $\xi(r_p,  \pi)$ is given in Figure \ref{zehavi}. This figure also exhibits another common distortion of the correlation function because of large scale coherent motions of galaxies around clusters and superclusters of galaxies leading to a compression of the correlation function in the perpendicular direction\cite{12}. Significant cosmological information can be obtained from measuring the anisotropy in $\xi(r_p,\pi)$ and comparing it to models for these redshift distortions\cite{13}.

I have focused here on the spatial two--point autocorrelation function, but there are many other correlation functions in operation. For example, one can compute the angular correlation function using just the positions of the galaxies in the plane of the sky. This has the disadvantage of diluting the true 3-D clustering signal as one is integrating over the radial distances. Likewise, one can also define the 3--point correlation function (both spatially and angularly) as the probability of finding a triplet of galaxies as a function of their triangular configuration. Recently there has been significant interest in computing such 3--point correlation function for galaxies because of the existence of large redshift surveys, like the Sloan Digital Sky Survey (SDSS; see Section \ref{obs}), and the availability of the required computational resources\cite{14}. Such 3--point functions can provide important information on the biasing of galaxies and how they populate their dark matter halos (see Section \ref{biasing})

\section{Clustering Measurements}
\label{obs}

\subsection{Power Spectrum of SDSS Galaxies}

There is a rich history of clustering measurements from galaxy surveys. From the pioneering measurements using the Shane--Wirtanen counts by Jim Peebles and collaborators to new digital catalogues like the Sloan Digital Sky Survey. In between, there have been important surveys like the APM and EDSGC angular galaxy catalogues based on digitized photographic Schmidt plates, the CfA Redshift Survey, based on the UGC and Zwicky catalogues, the Las Campanas Redshift Survey, the Durham/UKST Survey, the CNOC surveys and, more recently, the 2MASS survey and the 2dF Galaxy Redshift Survey (2dFGRS)\cite{15}.

I focus here on the Sloan Digital Sky Survey\cite{16} (SDSS) as it is currently the largest survey of the northern hemisphere in terms of numbers of galaxies and volume of the Universe sampled. In 2006, the SDSS published its fifth data release (DR5) which included redshifts for 674,749 galaxies as well as detections of many millions of fainter galaxies in the photometric sample. The SDSS will continue until July 2008, when a series of new surveys, using the same telescope and similar instrumentation, will begin.

\begin{figure*}[t]
 \includegraphics[width = 0.9\textwidth]{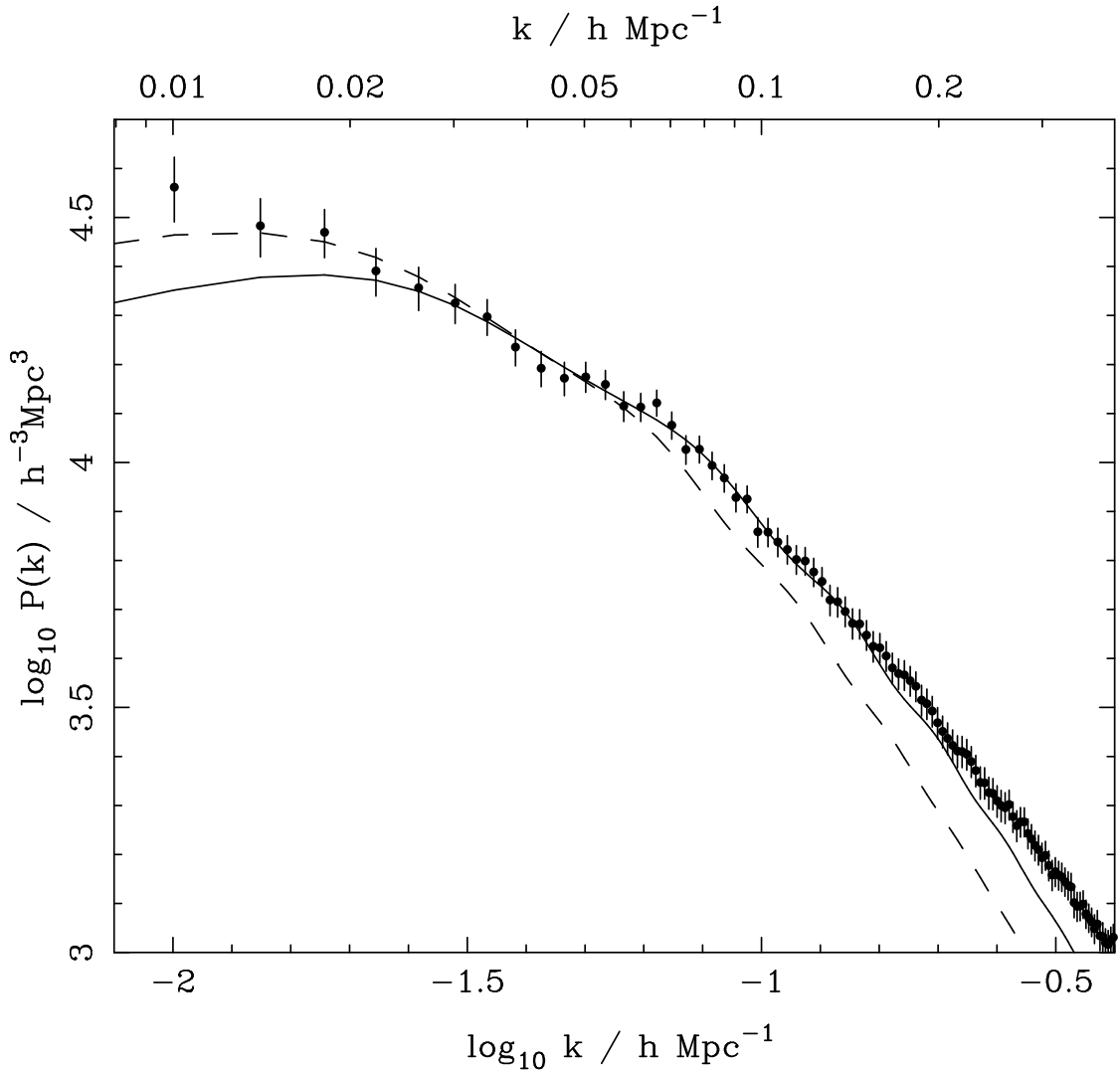}

\caption{The power spectrum of galaxy clustering from the SDSS DR5 data-set taken from \cite{17}. The dashed line shows the best fit Cold Dark Matter (CDM) model over the range $0.01<k<0.06h$ Mpc$^{-1}$ and has a value of $\Omega_m=0.22\pm0.04$. The solid line is the best fit CDM model over the range $0.01<k<0.15h$ Mpc$^{-1}$ and gives a value of $\Omega_m=0.32\pm0.01$. Therefore, the two models are inconsistent at $\simeq2\sigma$ and demonstrates that the SDSS $P(k)$ prefers a lower density Universe.}
\label{percival}
\end{figure*}

In Figure \ref{percival}, I show the power spectrum of galaxy clustering from the SDSS DR5 data\cite{17}, which raises two important issues. First, the $P(k)$ on large scales (small $k$) is still rising and has not turned over as show in Figure 3. This clearly disfavors the $\Omega_m=1$ model in Figure \ref{hutsi}, which begin to decline on scales of $k < 0.07h{\rm Mpc^{-1}}$. The constraints on $\Omega_m$ are even stronger than this, as illustrated in Figure \ref{percival}, where the data is best fit on large scales wit $\Omega_m = 0.22\pm 0.04$. It is therefore important that we push measurements of $P(k)$ to larger scales as this will improve the constraints on $\Omega_m$, especially if the expected decline in $P(k)$ is eventually seen. Such measurements however, will becomes increasingly difficult because of systematics issues associated with uncertainties in the galactic reddening corrections and the absolute photometric calibration across the survey (see Figure 18 of \cite{17}). 

The full SDSS galaxy redshift survey when completed in 2008 should provide an important insight on the large scale behaviour of $P(k)$. Likewise, photometric redshift surveys of galaxies can also be used to probe $P(k)$ on the largest possible scales, and recent results\cite{18} suggest significant large scale power in the power spectrum that could push $\Omega_m$ below 0.2. 

Secondly, the shape of the best fit $\Omega_m = 0.22$ model in Figure \ref{percival} is noticeably different from the data on scales $k>0.06h$ Mpc$^{-1}$. One would expect some differences on small scales because the model does not contain any information on the non--linear effects of gravity, which boost the observed clustering of matter relatively to the linear prediction. In addition, cosmologists must also deal with the annoyance of  ``galaxy bias", which is the relationship between the distribution of the underlying dark matter (which can not be seen directly) and the distribution of galaxies that can be seen in galaxy surveys like the SDSS. Naively, one assumes this relationship is a constant, e.g., $\delta_{galaxies} = b\delta_{darkmatter}$, which leads to $P_{galaxies}(k) = b^2 P_{darkmatter}(k)$. As can be seen in Figure  \ref{percival}, a simple constant offset between the galaxies (data points) and the dark matter predictions (lines) does not work on all scales as the shapes are different. Therefore, $b$ does appear to be scale--dependent and should be written as $b(k)$. One can now ask if $b$ is also a function of other parameters and if so, what does this mean for the use of $P_{galaxies}(k)$ as a precision cosmological tool. 

\subsection{Galaxy Biasing}
\label{biasing}

\begin{figure}[t]
\includegraphics[width = 0.8\textwidth]{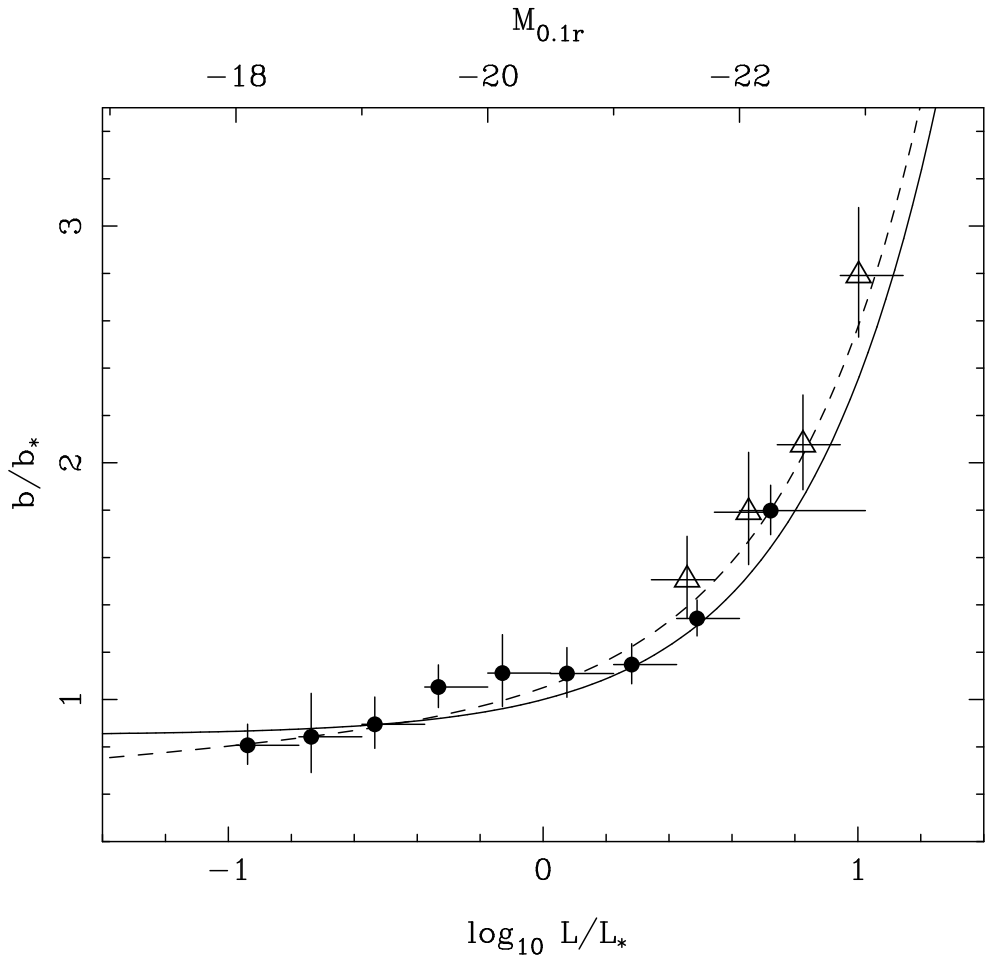}
  \caption{The bias values ($b$) for both the SDSS main galaxies (solid circles with
    one $\sigma$ error bars) and Luminous Red Galaxies (LRGs; open triangles with one $\sigma$ error bars) as a function of luminosity (or absolute magnitude, ${\rm M}_{^{0.1}r}$). The horizontal error bars
    show the range of luminosities in each bin. The solid line is the relation $b/b_*=0.85+0.15L/L_*$\cite{17}, while the
    dashed line is Eqn. \ref{biasform}. The latter is better for LRGs. This figure is taken from \cite{17}.\label{bias}}
\end{figure}

It has been known for some time that $b$ is a function of galaxy luminosity, with more luminous galaxies having higher $b$ values. Likewise, because of the color--magnitude relationship, redder (older) galaxies tend to be more biased as well, especially for the most luminous red galaxies (LRGs) lying at the centres of clusters of galaxies. Luckily, as shown in Figure \ref{bias}, the change in $b$ with colour and luminosity can be modeled\cite{19} and thus compensated for during any analysis of the $P_{galaxies}(k)$, e.g., the relative bias of galaxies is well fit by 

\begin{equation}
b/b_*=0.85+0.15L/L_*+0.04({\rm M}_*-{\rm M}_{^{0.1}r}),
\label{biasform}
\end{equation}

\noindent where $b_*$ is the bias of $L_*$ (${\rm M_*}$) galaxies and ${\rm M}_{^{0.1}r}$ is the absolute magnitude in a redshifted $z=0.1$ SDSS $r$-band filter. This latter term is a correction for the colour\cite{18} as well as the luminosity ($L$) of the galaxy. In this way, the $P_{galaxies}(k)$ of any galaxy sample (as a function of $L$ and colour) can be re-normalized and should agree. A key question is whether Eqn. \ref{biasform} is also a function of $k$, i.e., takes different functional forms at different scales? On large scales, where most of the cosmological information resides, the present data is insufficient to address this question. 

\begin{figure}[t]
\includegraphics[width = 0.8\textwidth]{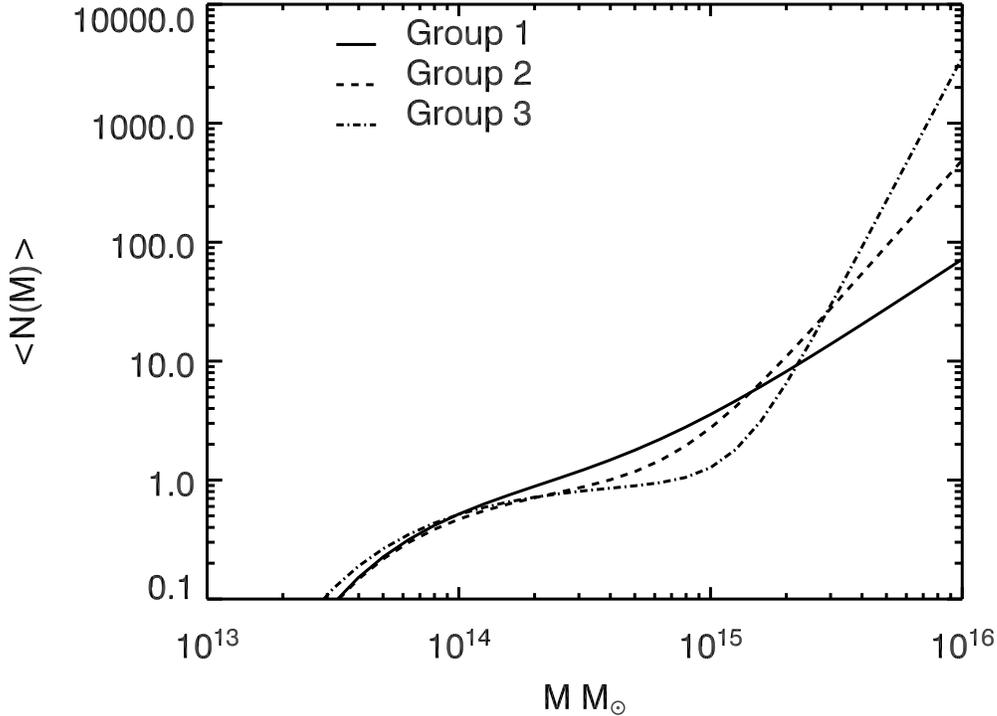}
\caption{Three examples of the Halo Occupation Distribution (HOD, Eqns. 6-8), which give the mean number of galaxies as a function of the halo mass ($\left < N(M) \right >$).  All three have the same number density of galaxies and two--point autocorrelation functions, but differ when studying the 3--point correlation function.  The figure is taken from \cite{21}.
\label{hod}}
\end{figure}

Recently, the issue of bias, and how it may depend on scale and galaxy properties, has been encapsulated in the ``halo model"\cite{20}. Here, the distribution of galaxies and dark matter are modeled as a distribution of distinct clumps, or ``halos", of dark matter with a Halo Occupation Distribution (HOD) explaining how the galaxies are located within these halos. A common parameterization of the HOD is
\begin{eqnarray}
 \left < N_{cent}(M)\right > & = & \mbox{exp}\left(- \frac { M_{min} }{M} 
 \right),
 \label{Ncent} \\
 \left < N_{sat}(M)\right > & = & \mbox{exp}\left(- \frac { M_{min} }{M}
 \right) \left( \frac{M}{M_1} \right) ^ \alpha ,
 \label{Nsat} \\
 \left < N(M) \right> & = & \left<N_{cent}(M)\right> + \left<N_{sat}(M)
 \right> .
 \label{NM}
\end{eqnarray}
The terms $cent$ and $sat$ refer to the central galaxy, the single galaxy at the centre of the dark matter halo, and satellite galaxies, which are located elsewhere in the halo. If a halo only has one galaxy, it is a defined to be the central galaxy. $M_{min}$ is the minimum mass below which it is exponentially unlikely to host even one (central) galaxy, while $M_1$ denotes the mass required to 
host at least one satellite galaxy. Finally, $\alpha$ is the slope of the power--law relation for adding more satellites as a function of halo mass.  Figure \ref{hod} shows three examples of this HOD for different values of  $M_{min}$, $M_1$ and $\alpha$. Interestingly, all three of these HODs produce the same mean space density of LRGs as well as the same two--point correlation functions. Only by studying the 3--point correlation function can the degeneracy between these HOD parameters be broken\cite{21}.

\subsection{BAO Observations}

\begin{figure}[t]
\includegraphics[width = 0.8\textwidth]{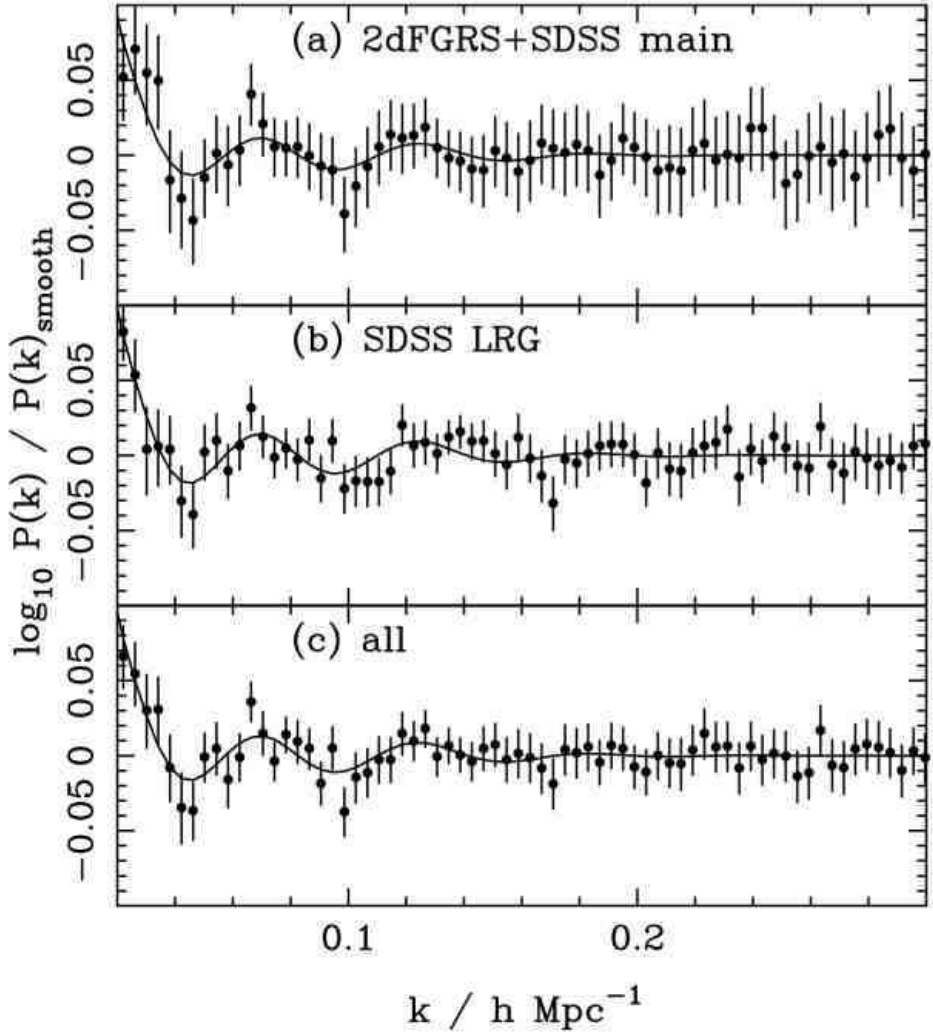}
\caption{
Recent measurements of the BAO taken from \cite{24} using a combination of data from the 2dFGRS and the SDSS. The smooth component of the power spectrum (as seen in Figures 3 and \ref{percival}) has been subtracted thus leaving only the BAO signal. The top panel is the combination of the 2dFGRS sample (143k unique galaxies not in the SDSS area) and the main galaxy redshift survey of the SDSS (466k galaxies). The middle panel is just the SDSS Luminous Red Galaxy sample (78k galaxies), which probes higher redshifts than the SDSS main galaxy sample and 2dFGRS. The bottom panel is the combination of all data. The solid lines are the best fits to the BAO signal using a model for the BAO. The one sigma errors are shown.
\label{baofig}}
\end{figure}

This decade has seen significant advances in the detection and measurement of the BAO in the galaxy distribution (Section \ref{theory}). The first claims of the detection of the BAO came in 2001\cite{22}, followed in 2005 by definitive results by several authors\cite{23}. In Figure \ref{baofig}, I show the latest BAO measurements\cite{24} using all available data from the SDSS and 2dFGRS. As can be seen, the BAO are now clearly detected\cite{25} in several different samples of galaxies spread over most of the extragalactic sky. 

We have now moved from the ``detection" phase of the BAO into the ``measurement" phase, and are now using the BAO as a ``standard ruler" in the Universe, as they represent a fixed metric scale imprinted in the distribution of matter at recombination (See Section \ref{theory}). Quite simply, if one can measure the BAO at different redshifts, then one can measure the rate of change of distance in the Universe and thus the acceleration in the expansion of the Universe. In reality, the BAO measurements provide a combination of the angular-diameter distance, because of the BAO signal in the plane of the sky, and the Hubble parameter, from the BAO along the line--of--sight. Therefore, the BAO measurements constrain a ``volume--averaged" distance ($D_v(z)$) as the galaxy data is usually analysed in thick redshift shells, e.g.,

\begin{equation}
D_v (z) = \left[ \frac{(1+z)^2 c z D_A(z)^2}{H(z)}\right]^{\frac{1}{3}}  
\label{distvol}
\end{equation}

\noindent where $D_A(z)$ is the angular--diameter distance, as a function of redshift, and $H(z)$ is the Hubble parameter with redshift. In the future, with larger galaxy catalogues, it maybe possible to decouple this cosmological information ($D_A(z)$ and $H(z)$) and determine the BAO signal in both the transverse and line--of--sight directions as in the $\xi(r_p,\pi)$ in Figure \ref{zehavi}. In this case, one would see a circular ``BAO ridge'', in the correlation function, or ``BAO rings", in the power spectrum\cite{26}. 

Using the data in Figure \ref{baofig}, it is now possible to measure $D_v$ in two different redshift shells\cite{24} and compute the ratio of distances between these two shells based solely on the BAO signal, e.g.,  $D_v(0.35) / D_v(0.2)$, the ratio of $D_v$ at redshifts 0.2 and 0.35 (the median redshifts of the two shells shown in Figure 8). Using the SDSS and 2dFGRS data, we find $D_v(0.35) / D_v(0.2) = 1.812\pm0.060$, which is higher, by $2.4\sigma$, than the expected distance ratio for a flat $\Lambda$--dominated CDM model ($\Lambda$CDM) with $\Omega_m=0.25$ (which is $D_v(0.35) / D_v(0.2) = 1.67$). Systematic uncertainties associated with how the BAO signal is modeled and fitted, as well as the damping effects of non-linear structure formation, can change this distance ratio by $\sim1 \sigma$, so the discrepancy with  flat $\Lambda$CDM  could be decreased. 

At face value however, this measurement of the BAO distance ratio indicates there maybe more acceleration in the local Universe ($z<0.4$) than is presently expected given the combination of the higher redshift CMB and supernova data. This maybe an indication of intrinsic curvature in the Universe, or a dynamical $w(z)$: We will see. Otherwise, if we combine all the data together (CMB, BAO, supernovae), then the best fit cosmological parameters are in agreement with previous measurements, e.g., $\Omega_m=0.252\pm0.027$ and $w=-1.004\pm0.088$ (assuming flatness and a constant $w$). However, this does mask the underlying tension between the BAO and supernova distance measurements (at the $\simeq2\sigma$ level).  

\section{Integrated Sachs-Wolfe Effect}
\label{isw}

\subsection{Background}

In the previous section, I focused on 3-D galaxy maps of the Universe based on large redshift surveys. However, 2-D galaxy maps can also provide important information on the large scale structure (LSS) in the Universe, especially when combined with photometric redshift estimates (``photo-z's") based on the colours of galaxies. Such photo-z's are now reliable with a mean error on the true redshift of between $0.03$ and $0.06$ for LRGs (depending on the LRG brightness and redshift) over the redshift range $0.4<z<0.7$\cite{27}. These errors will get better as larger training samples are developed. 

One recent use of large 2-D galaxy maps has been the detection and measurement of the late--time Integrated Sachs-Wolfe (ISW) effect, which is the change in the energy of CMB photons as they travel to us from the surface of last scattering through time-evolving gravitational potential wells. Briefly, as a CMB photon falls into a large mass overdensity it experiences a gravitational blueshift, i.e., it gains energy. Likewise, as the photon leaves the overdensity, it gives up that extra energy as a gravitational redshift. If the potential well has not changed during the passage of the photon, then the two effects cancel and the photons' energy remains unchanged. However, if the potential well has evolved during the photon passage, then it will leave with either a boost in energy or a loss. Therefore, the late--time ISW effect is the integrated effect of all these blue and redshifts along the line--of--sight, and is only considered in the linear regime of the power spectrum of density fluctuations (on large scales). The Rees--Sciama effect\cite{28} is the analogous effect in the non--linear regime (on small scales), while bulk motions of masses along the line--of--sight will also add to the effect\cite{29}. I ignore these small scale effects herein, but they could become important in the future as the resolution of CMB maps improves.  

In a matter--dominated Universe, potential wells remain constant with respect to co--moving coordinates, so we would not expect to see an ISW effect. However, if the Universe becomes dominated by dark energy (or spatial curvature if we assume $\Omega_m<1$), then the large scale gravitational potentials will evolve and we expect to see an ISW effect. 

It is hard to measure the ISW effect from the CMB data alone\cite{30}, as it is a sub--dominant signal on large scales in the CMB angular power spectrum. Alternatively, one can cross--correlate the CMB maps with tracers of the gravitational potential wells and so separate the ISW signal from the intrinsic fluctuations in the CMB\cite{31}. As dark energy only dominates the energy density of the Universe at relatively late epochs (below $z\sim0.5$), then we only need to correlate ``local" galaxy maps with the CMB to detect this effect (see below). 

Unfortunately, the ISW effect will not provide precision measurements of the cosmological parameters because it is inherently noisy. On large scales, the CMB primary fluctuations dominate the signal in addition to the problem of cosmic variance. Predictions of the ISW effect suggest that at best one can hope for a $7.5\sigma$ detection of the signal, or more realistically, a $\simeq5\sigma$ detection given an all--sky galaxy map with $10^7$ galaxies out to $z\sim1$\cite{32}. However, such predictions are sensitive to the assumed density of dark matter and therefore, higher detection significances could be used to place interesting lower limits of $\Omega_m$. 

Regardless of this fundamental limit on the detectability of the ISW effect, it does provide complementary information to other cosmological probes and can still be competitive. For example, future measurements of the ISW effect using the cross--correlation of the LSST galaxy data (www.lsst.org) and Planck CMB data will provide competitive constraints on dynamical dark energy models (models with a varying $w(z)$), compared to the next generation of supernova and BAO experiments\cite{33}. Likewise, the ISW provides an opportunity to constrain the sound speed of dark energy\cite{34}. 

\begin{figure}[t]
\includegraphics[width = 0.8\textwidth]{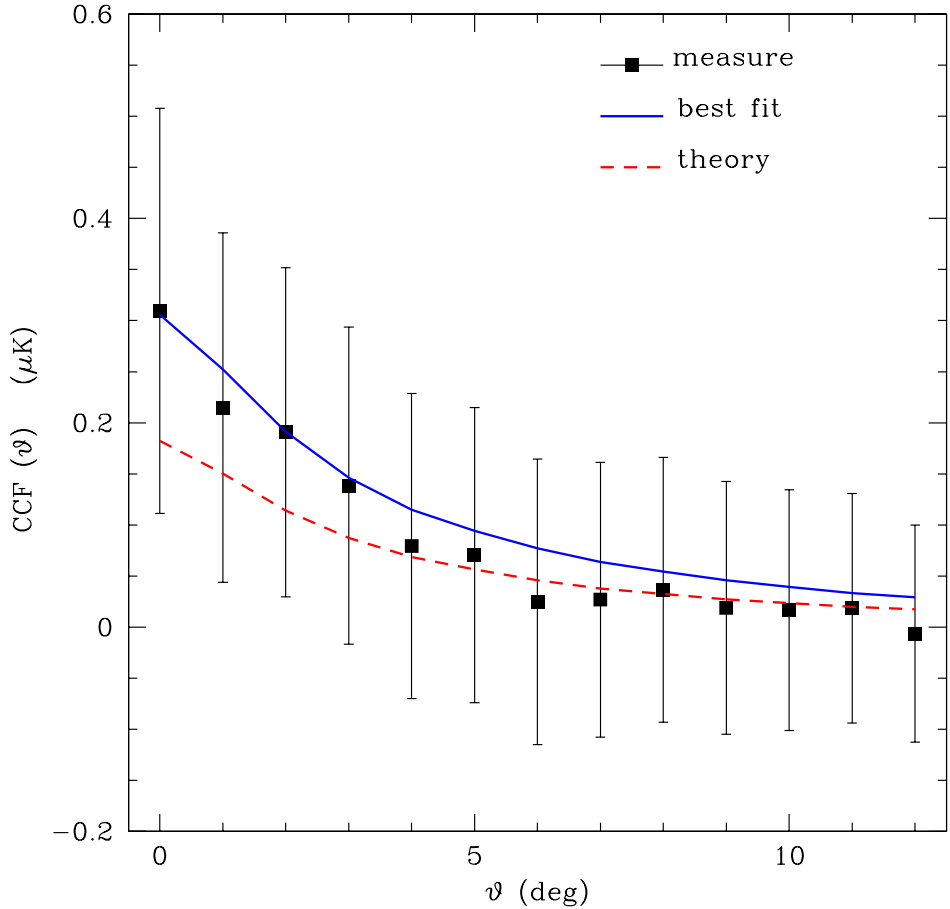}
\caption{The angular cross--correlation function of photometrically classified quasars from the SDSS (at $z>1$) and the WMAP CMB data. The square points are the observed correlation, while the solid line is the best fit  $ \Lambda $CDM theoretical model and the dashed line is the prediction for the WMAP3 best fit model with a bias of $ b = 2.3 $ for the quasars. The points and error bars are highly correlated, e.g., the typical level of correlation between two neighbouring bins is $ \sim 95 \% $. Figure taken from \cite{37}}\label {tom}\
\end{figure}

\subsection{ISW Measurements}

This decade has seen significant progress in detecting the late-time ISW effect because of the availability of the WMAP CMB data and new large--area surveys like the SDSS. The ISW effect has now been seen using an array of different galaxy maps (Xray, optical, radio) and at several different redshifts\cite{35}. There is some disagreement about the significance level of several of these ISW detections, but overall most authors find a positive correlation between the CMB and LSS at the $2$ to $3\sigma$ level. The reason for these discrepancies is the difficulty in assessing the likelihood of any observed correlation because of cosmic variance. This can either be assessed using jack--knife (JK) error estimates, which are subjective in their choice of the size and number of the JK sub--regions, or Monte Carlo errors using thousands of fake CMB maps that must be generated using the observed CMB angular power spectrum\cite{36}. The latter is computationally expensive and present analyses have been restricted to cross-correlating the observed galaxy maps against $\simeq5000$ fake CMB maps. The significance of detection is also complicated by the fact that many of the different galaxy maps  are themselves correlated and therefore, to add all of the detections together, one needs to account for the cross-correlations between the different LSS tracers. Future work on this should bring together all these ISW detections, thus providing the best, single dataset for constraining cosmological parameters.

One of the benefits of ISW measurements is the fact they can be carried out at any redshift where sufficient tracers of the LSS can be obtained. Crudely, any catalogue of extragalactic objects with a surface density of $>50$--$100$ per square degree, and covering several thousands of square degrees of sky, can be used to measure the ISW effect. In Figure \ref{tom}, we show an example of this potential, which is the  cross--correlation of SDSS photometrically--selected quasars at $z\sim1.5$ with the WMAP data\cite{37}. This figure shows a positive correlation at the $\simeq2.5\sigma$ significance (compared to no correlation) consistent with that expected for $ \Lambda $CDM. 

\begin{figure}[t]
\includegraphics[width = 0.8\textwidth]{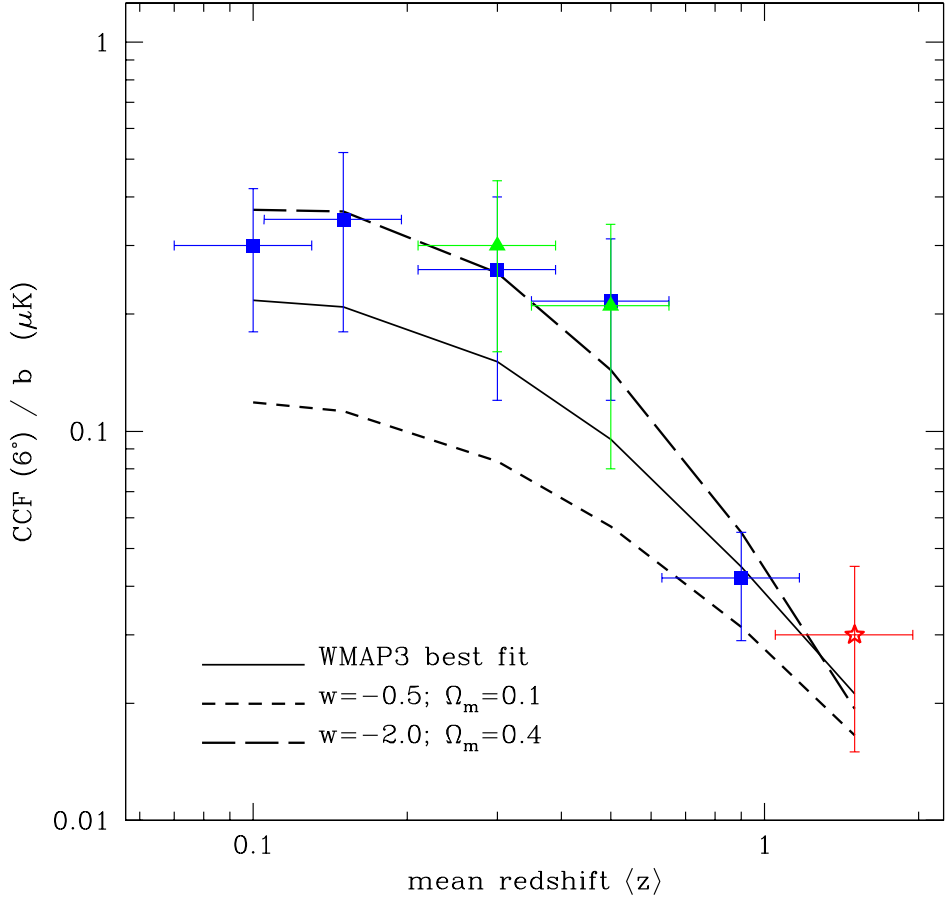}
\caption{
Summary of the detections of the ISW effect. The blue (squared) and green (triangular) points are cross--correlations for a number of galaxy catalogues (2MASS, APM, SDSS, SDSS high-z, NVSS+HEAO, see \cite{35}. The red (star) point is the SDSS quasars as discussed in the text\cite{37}. The lines are theoretical expectations for WMAP3 best fit $\Lambda$ model (solid), and two models with $ w = -2 $ (long dashed) and  $ w = -0.5 $ (short dashed) respectively. Figure taken fropm \cite{37}}\label {tom2}
\end{figure}

At face value, this result shows evidence for dark energy even at these high redshifts ($z>1$) and is the earliest epoch that such measurements have been achieved\cite{38}. The ISW effect allows us to probe the behaviour of dark energy at these high redshifts independent of the low redshift density of dark energy. The same is not true for geometrical tests of the Universe, like supernovae and BAO, as they are integrated quantities along the line--of--sight and thus measure average values for the density of dark energy out to high redshift. As our tracers are at $z>1$, we are measuring the density of dark energy at these epochs, e.g., assuming a cosmological constant, then Figure 9 tells us that $\Omega_{de} (z=1.5) = 0.16$. The expected drop in the density of dark energy with redshift is shown in Figure \ref{tom2} which presents several ISW measurements as a function of redshift compared to that expected for the standard $\Lambda $CDM model (best fit to the WMAP3 data). It is reassuring that there is good agreement between the two, thus indicating that more complex equations of state for dark energy are not presently needed. This figure again reinforces the claim  that the ISW effect will provide competitive constraints on dynamical dark energy models in the future.

In addition to constraining dynamical dark energy models, several authors have highlighted the power of the ISW effect for testing theories of modified gravity. This is primarily because the ISW effect is sensitive to the rate of structure formation in the linear regime, which is related to the strength of gravity on these large scales. One popular model for modified gravity is the Dvali--Gabadadze--Porrati (DGP) braneworld model\cite{39}, where our Universe is embedded in a higher--dimensional space and on large scales, gravity becomes a 5--D force rather than 4--D on smaller scales. This appears as a weakening of the gravitational force and thus would appear as an acceleration in the 4--D Universe on large scales. This manifests itself as a different decay rate for the large scale potential wells in the DGP model compared to a dark energy model, even if the two models have the same geometry. In this way, the ISW effect is complementary in its ability to probe these different physical models for the Universe, while geometrical measurements (like supernovae and BAO) would give the same result. Measuring the decay rate of the potential wells using high redshift tracers of the LSS is the way forward and could be a key test for these modified gravity models\cite{40}.

Finally, as measurements of the ISW effect push to higher redshift, the effects of gravitational lensing, through the magnification bias, and re-ionization\cite{41} in the Universe must be taken into account. This could be a curse or a blessing as it demonstrates that galaxy--CMB cross--correlations potentially contain lots of interesting information at high redshift.

\section*{Acknowledgments}

The work presented in this review was taken from collaborative research with many colleagues including Will Percival, Ryan Scranton, Tommaso Giannantonio, Robert Crittenden, and many others in the SDSS collaboration. I thank these authors for allowing me to present their work here. I thank Martin White for his informative BAO webpages, which I heavily used in Section \ref{theory}. I also thank Chris Clarkson for helpful conversations about the particle horizon scale at matter--radiation equality, and Gert H{\"u}tsi and Idit Zehavi for reading earlier drafts of this review.  Finally, I thank the referees for their comments, which improved this review.

Funding for the SDSS and SDSS-II has been provided by the Alfred P. Sloan Foundation, the Participating Institutions, the National Science Foundation, the U.S. Department of Energy, the National Aeronautics and Space Administration, the Japanese Monbukagakusho, the Max Planck Society, and the Higher Education Funding Council for England. The SDSS Web Site is http://www.sdss.org/.

The SDSS is managed by the Astrophysical Research Consortium for the Participating Institutions. The Participating Institutions are the American Museum of Natural History, Astrophysical Institute Potsdam, University of Basel, University of Cambridge, Case Western Reserve University, University of Chicago, Drexel University, Fermilab, the Institute for Advanced Study, the Japan Participation Group, Johns Hopkins University, the Joint Institute for Nuclear Astrophysics, the Kavli Institute for Particle Astrophysics and Cosmology, the Korean Scientist Group, the Chinese Academy of Sciences (LAMOST), Los Alamos National Laboratory, the Max-Planck-Institute for Astronomy (MPIA), the Max-Planck-Institute for Astrophysics (MPA), New Mexico State University, Ohio State University, University of Pittsburgh, University of Portsmouth, Princeton University, the United States Naval Observatory, and the University of Washington.

\end{document}